\documentclass[prd,onecolumn,preprintnumbers,nofootinbib]{revtex4}
\usepackage{amsfonts,amsmath,amssymb}
\usepackage{graphicx}
\usepackage{color}
\usepackage{natbib}
\usepackage{siunitx}
\usepackage{threeparttable}

\usepackage[plainpages=false, colorlinks=true, anchorcolor=blue, linkcolor=blue, citecolor=blue, bookmarks=false]{hyperref}
\newcommand{\rthis}[1]{\textcolor{black}{#1}}

\usepackage{natbib}

\begin{document}
%\markboth{Authors' Names}

%%%%%%%% Journals %%%%%%%%%%%%%%

%%%%%%%%%%%%%%%%%% TITLE %%%%%%%%%%%%%%%%%%%%%%%%%%%%%%%%%%%% 
\title{A meta-analysis of  distance measurements   to M87}
\author{Gunasekar \surname{Ramakrishnan}$^1$}  \altaffiliation{E-mail: ramkrish1401@gmail.com}
\author{Shantanu  \surname{Desai}$^2$} \altaffiliation{E-mail: shntn05@gmail.com}

\affiliation{$^{1}$Department of Physics, Muthurangam Government Arts College, Vellore, Tamil Nadu - 632002,  India}
\affiliation{$^{2}$Department of Physics, Indian Institute of Technology, Hyderabad, Telangana - 502285, India}

\begin{abstract}
We obtain  the  median, arithmetic mean,  and  the  weighted mean-based central estimates for the distance to M87 using   all the measurements  collated  in ~\cite{DeGrijs19}. 
We then reconstruct the error distribution for the  residuals of the combined  measurements and also splitting  them based on the tracers used.   We  then checked for  consistency with a Gaussian distribution and  other symmetric distributions such as Cauchy, Laplacian, and Students-$t$ distributions. We find that when we analyze the combined data, the weighted mean-based estimates show a poor agreement with the Gaussian distribution, indicating that there are unaccounted systematic errors in some of the measurements. Therefore, the median-based estimate for the distance to M87 would be the most robust. This median-based distance modulus to M87  is given by  $31.08 \pm 0.09$ mag  and $31.07 \pm 0.09$ mag, with and without  considering measurements categorized as ``averages'', respectively. This estimate agrees with the corresponding value obtained in ~\cite{DeGrijs19} to within $1\sigma$.    

%\pacs{97.60.Jd, 04.80.Cc, 95.30.Sf}
\end{abstract}

\maketitle

%%%%%%%%%%%%%%%%%%%%%%%%%%%%%%%%%%%%%%%%%%%%%%%%%%%%%%%%%%%
%%%%%%%%%%%%%%%%%%%%%%%%%%%%%%%%%%%%%%%%%%%%%%%%%%%%%%%%%%%
%%%%%%%%%%%%%%%%%%%%%%%%%%%%%%%%%%%%%%%%%%%%%%%%%%%%%%%%%%%
%\noindent {\it Introduction.---}
\section{Introduction}
The VIRGO cluster and its giant elliptical galaxy M87 is an important anchor for the distance estimates to more distant astronomical objects such as the Fornax and Coma cluster. Therefore, De Grijs et al~\cite{DeGrijs19} (D20 hereafter) have done an extensive data mining of all distance measurements  to M87/Virgo cluster and compiled a database of  213 distances. D20 grouped these measurements into five categories, depending on the method used. They obtained a distance modulus of $(m-M)=31.03 \pm 0.14$ mag corresponding to a distance measurement 
of $16.07 \pm 1.03$~Mpc. This central estimate was obtained using the weighted mean.

%The central estimate of the neutron lifetime  mentioned in PDG as well as all other works, which analyze this discrepancy has been obtained from a weighted average of all the measurements. \rthis {The central estimate of a quantity using weighted measurements makes the following  main assumptions~\cite{Gott}: (i) individual data points are statistically independent and contain no systematic effects ; (ii) the errors are Gaussianly distributed. If any of the measurements contain catastrophic outliers or unaccounted systematic effects, then the second assumption is automatically violated. In that case, the weighted mean can produce extremely biased results. On the other hand, median statistics does not incorporate the individual measurements errors, and hence is unaffected by the presence of a few outliers. Secondly, even if the errors are not correctly estimated, as shown using simulations of Zeldovich's thought experiment involving watches~\cite{Bethapudi}, median estimate gives a more robust estimate.  Even if a dataset is drawn from a distribution with infinite variance such as Cauchy distribution, the median is a more robust central estimate~\cite{Gott}. Many additional pitfalls in using the weighted mean as a central estimate, and how using the median value ameliorates these problems can be found in Refs.~\cite{Gott, Bethapudi} and references therein. The only assumption used for median statistic based estimate is that the measurements are independent and free of systematic errors.}

A large number of studies (mainly by Ratra and collaborators) have shown that the error distributions for a whole suite of  astrophysical and cosmological measurements  are not consistent with a Gaussian distribution~\citep{Gott,Podariu01,Chen03,Chen,Ratra03,Ratra15,RatraLMC,Crandall,Bethapudi,Rajan,RatraD2,Zhang18,Ratratheta,Yu20,Zhang22}. Some examples  are measurements of $H_0$~\cite{Gott,Ratra03,Zhang18,Bethapudi}, Lithium-7 measurements~\cite{Ratra15} (see also ~\cite{Zhang}), distance to LMC ~\cite{RatraLMC}, distance to the galactic center~\cite{RatraGC}, Deuterium abundance~\cite{RatraD2}, individual data points used to measure Hubble constant~\cite{Thakur}, \rthis{CMB anisotropy detections~\cite{Podariu01}}, etc. A similar analysis has also been done for particle physics data~\cite{Rajan2,Ashish,Zhang22} and Newton's constant~\cite{Bethapudi,Desai16,Bhagvati}. Such studies have also been done in the field of medicine and psychology~\cite{Bailey}.
For the aforementioned  datasets, the above works  fit the error residuals to multiple  probability distributions.  From most of   the above studies,  it was inferred  that the error distribution for the analyzed  datasets is not Gaussian. Therefore, it was  argued that median statistics should be used for the central estimate, instead of the often used weighted mean~\citep{Gott,Bethapudi}.  Therefore, median statistics  has   been used to obtain central estimates of some of these quantities such as Hubble Constant~\cite{Gott,Chen,Bethapudi}, Newton's Gravitational Constant~\cite{Bethapudi}, neutron lifetime~\cite{Rajan2}, mean matter density~\cite{Chen03}, and  cosmological parameters~\cite{Crandall}.

Given the important astrophysical and cosmological implications of the distance to the Virgo cluster from Hubble constant~\cite{Kim20} to imaging of the black hole in M87~\cite{EHT}, estimating the distance to Fornax and Coma clusters, it is paramount to  get a more robust estimate to M87. For this purpose,  we revisit the issue of checking for non-Gaussianity of the error residuals   using the measurements compiled in D20.
The manuscript is structured as follows. The dataset used for our analysis is described in Sec.~\ref{sec:dataset}. Our analysis procedure  is described in Sec.~\ref{sec:analysis}.  Our results can be found in Sec.~\ref{sec:prob}. Our conclusions are descibed in   Sec.~\ref{sec:conclusions}.

\section{Distance measurements to M87/Virgo cluster}
\label{sec:dataset}
We briefly review the data used for this analysis . More details can be found in D20 and references therein. D20 perused the NASA/ADS database (up to Sept 3, 2019) using the keywords `M87' and obtained 213 independent distance estimations starting from Hubble 1929 measurement~\cite{Hubble29} to Hartke's 2017 analysis~\cite{Hartke17}.  Only those measurements associated with the  M87 subcluster or centered around M87 were used. Their final catalog consists of 213 measurements out of which 173 have error bars. These have been collated at \url{http://astro-expat.info/Data/pubbias.html}. These measurements have been divided into 15 tracers. Out of these,  one set of tracers consists of ``Averages'', which is a collation of 21 papers, with each paper containing averages of heterogeneous measurements of different types. Another category is called ``Other methods'', which consists of 15 independent measurements without any proper classification. These range from unspecified methods to techniques, which are independent of any distance ladder and purely based on Physics principles, such as the Sunyaev-Zeldovich observation of the VIRGO cluster from Planck~\cite{Planck16}.  Among these 15 tracers, eight tracers have more than 10 measurements. Finally, we note that D20 has tabulated the distance measurements in terms of the distance modulus which is measured in  AB magnitudes. 
We note that the published distances, whenever applicable have been homogenized to conform with the distance modulus to LMC, $(m-M)^{LMC} = 18.49$ mag~\cite{LMC}. One can trivially convert the measurements of the  distance moduli into physical units of distance. \rthis{We should  note that we are using using the true distance moduli, i.e. the unreddened distance moduli.} 
 The recommended best-fit value of the distance to M87 obtained in D20 using the  weighted mean  is given by $(m-M)=31.03 \pm 0.14$ mag~\cite{DeGrijs19}.

%They also wanted to see if there is evidence for publication bias or ``bandwagon'' effect. 

\section{Analysis}
\label{sec:analysis}

The first step in assessing the Gaussianity of the error measurements of a dataset is to obtain the  central estimate. For this purpose, we use all the measurements collated in D20. We obtained the central estimate using median, weighted, and arithmetic mean.  The median estimate  $(m-M)_{med}$  corresponds to the 50\% percentile value. The standard deviation of the median depends upon the distribution  from where it is sampled from. Multiple methods have been proposed  to estimate the sample variance of the median~\cite{Woodruff,Maritz,Price}. 
Although, in our previous works we have used the prescription in ~\cite{Gott} to get the error estimate, we use the following equation to get the uncertainty in the median estimate~\cite{astroML}:
\begin{equation}
\sigma_{med} = \sigma\times\sqrt{\pi/2N},
\end{equation}
where $N$ is the number of data points and \rthis{$\sigma$ is the sample standard deviation. Note however that  the expression for $\sigma_{med}$ is mainly valid for Gaussian distributions as opposed to the method proposed in ~\cite{Gott}.}

The weighted mean   ($(m-M)_{wm}$) using the observed distance modulus measurements ($(m-M)_i$) is given by~\cite{Bevington}:
\begin{equation}
(m-M)_{wm} = \frac{\sum \limits_{i=1}^N \ (m-M)_i/\sigma_i^2}{\sum \limits_{i=1}^N 1/\sigma_i^2},
\end{equation}
\noindent where  $\sigma_i$ denotes the total error in each measurement. The error in the  weighted mean is given by: 
\begin{equation}
\sigma_{wm} = \frac{1}{\sqrt{\sum \limits_{i=1}^N 1/\sigma_i^2}}.
\end{equation}
The arithmetic mean central estimate ($(m-M)_m$) is given by:
\begin{equation}
(m-M)_m = \frac{1}{N}{{\sum\limits_{i=1}^N}(m-M)_{i}},
\end{equation}
with the standard deviation given by:
\begin{equation}
\sigma_{m} = \sqrt{{\frac{1}{N^2}}{\sum\limits_{i=1}^N}{((m-M)_{i}-(m-M)_{m})^2}}.
\end{equation}

For any central estimate based on the median or arithmetic mean, we include all the tabulated measurements, irrespective of whether they are provided with error bars or not. For the weighted mean, we only include the measurements which have uncertainties. \rthis{Although, in principle one could also restrict the median or arithmetic mean based analysis to those measurements which only have error estimates, we decided to use the full dataset for computations which do not need the uncertainty estimates for increased statistics.}
 From the measurements in Table~\ref{tab:1}, the weighted mean estimate is found to be $(m-M)_{wm} = 31.11 \pm 0.008$,  whereas  the median estimate is equal to  $31.08 \pm 0.09$, and the arithmetic mean is equal to $30.97 \pm 0.07$. We also estimated the same after excluding the measurements tagged as  ``averages''. These values can be found in Table~\ref{tab:1}. Therefore the results are consistent with  each other to within 1$\sigma$. The central estimates are also consistent with the measurements in D20 to within $1\sigma$.

\begin{figure}
     \centering
        \includegraphics[width=0.7\textwidth]{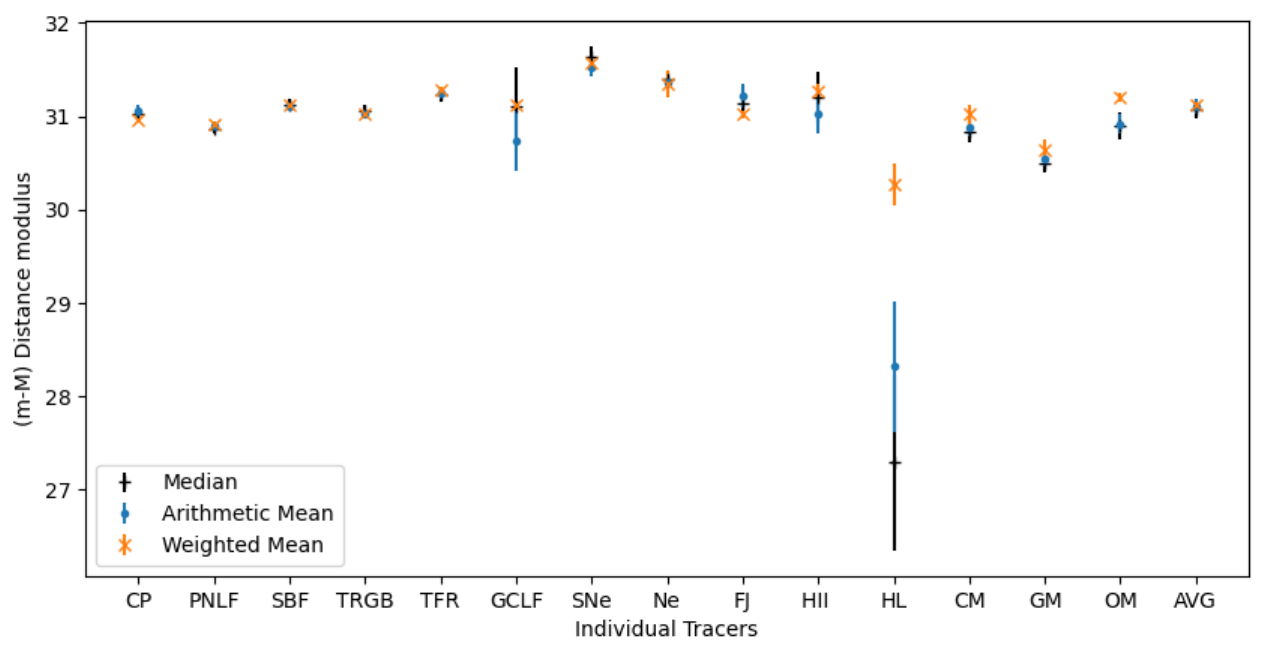}
         \label{fig:fig1}
            \caption{Central estimates of distance modulus to M87 for all the three methods groups according to category. Cepheids (CP), Planetary Nebula Luminosity Function (PNLF), Surface Brightness Variation (SBF), Tip of the Red-Gaint Branch (TRGB), Tully-Fisher Relation (TFR), Globular Cluster Luminosity Function (GCLF), Supernovae (SNe), Novae (Ne), Faber-Jackson relation (FJ), HII region (HII), Hubble Law (HL), Color-magnitude/luminosity (CM), Group membership (GM), Other methods (OM), Averages (AVG).}
 \label{fig:alpha}
\end{figure}

 We also obtained the arithmetic mean, weighted mean, and median for each of the different measurements grouped according to the tracers. These results  can be found in Table~\ref{tab:2}.  A graphical summary of the same can be found in Fig.~\ref{fig:alpha}.  We find that the measurements obtained based on Hubble's law have the largest error bars and are also discrepant with respect to the other measurements. It is also inconsistent with the D20 estimate at about 2.7$\sigma$  (arithmetic mean) to 3.8$\sigma$ (median estimate).
 
  We now check for the Gaussianity of the residuals using the combined dataset as well as using  the measurements grouped according to the tracer used. For the latter, we only consider the Gaussianity  as long as the number of independent measurements within each tracer is greater than 10. Such an analysis will guide us in choosing the most robust central estimate.

\subsection{Error Residuals}
\label{sec:analysisA}
After obtaining the  central estimate for the distance  $(m-M)_{CE}$ modulus to M87 using each of the aforementioned methods, we calculate the residual  error    as follows~\cite{RatraD2,RatraGC}:
\begin{equation}
N_{\sigma_i} =\frac{(m-M)_i-(m-M)_{CE}}{\sqrt{\sigma_i^2+\sigma_{CE}^2}} ,
\label{eq:nsigma}
\end{equation}
Eq.~\ref{eq:nsigma} is used for  $N_{\sigma_i}^{med}$, $N_{\sigma_i}^{m}$, $N_{\sigma_i}^{wm+}$, where $\sigma_{CE}$ denotes the error in the central estimate for each of the different methods,  and $\sigma_i$ is the error in the individual measurements. As in Refs.~\cite{RatraD2,RatraGC,Ratratheta}, we denote our error distribution for the median ($(m-M)_{med}$), arithmetic mean ($(m-M)_{m}$) and the  weighted mean ($(m-M)_{wm}$) calculated from Eq.~\ref{eq:nsigma} by $N_{\sigma_i}^{med}$, $N_{\sigma_i}^{m}$, and  $N_{\sigma_i}^{wm+}$, respectively.  When the  central  estimate is obtained from the weighted mean, one  should take into account the correlations and the modified version of the error distribution, which accounts for these correlations becomes~\cite{RatraGC}:
\begin{equation}
N_{\sigma_i}^{wm-} =\frac{(m-M)_i-(m-M)_{CE}}{\sqrt{\sigma_i^2-\sigma_{CE}^2}}
\end{equation}

Therefore the only difference between $N_{\sigma_i}^{wm-}$ and $N_{\sigma_i}^{wm+}$ is that the latter does not include correlations.
Each of the above  sets  of  $|N_{\sigma}|$ histograms is then symmetrized around zero. We now fit the symmetrized histogram distribution  of $|N_{\sigma_i}|$ to multiple probability distributions as described in the next section.

\section{Fits of residuals to probability distributions}
\label{sec:prob}
We now fit the symmetrized  $|N_{\sigma}|$   histograms  to a Gaussian  distribution as well as  other symmetric distributions, such as the Cauchy, Laplacian, and Student's $t$ distribution, to test the efficacy of the each of these distributions. We briefly recap the different distributions used to fit the data.
%This is  similar in spirit  to our previous work on galactic rotation velocities~\cite{Rajan} or  neutron lifetime measurements~\cite{Rajan2}, which in turn followed the prescription in numerous works  by Ratra et al.  

The Gaussian distribution  has  a mean of zero  and  standard deviation of unity:
\begin{equation}
P(N) = \frac{1}{\sqrt{2\pi}}\exp(-|N|^2/2).
\label{eq:gauss}
\end{equation}

The second distribution we consider is the Laplacian distribution:
\begin{equation}
P(N) = \frac{1}{2}\exp(-|N|)
\label{eq:laplace}
\end{equation}

The third distribution, which we shall use is the Cauchy or Lorentzian distribution. It  can be described by:
\begin{equation}
P(N) = \frac{1}{\pi(1+|N|^2)}
\label{eq:cauchy}
\end{equation}

Finally, the last distribution considered is the Student's-$t$ distribution, given by $n$ (or  ``degrees of freedom'') and is given by~\cite{astroML}:
\begin{equation}
P(N) = \frac{\Gamma[(n+1)/2]}{\sqrt{\pi n}\Gamma(n/2)(1+|N|^2/n)^{(n+1)/2}}
\label{eq:student}
\end{equation}
\noindent For $n=1$, the Student's-$t$ distribution reduces to  the  Cauchy distribution, and is same as the  Gaussian distribution for $n=\infty$. Similar to ~\cite{Rajan2}, we find the optimum value of $n$  in the range from 2 to 2000. 
We also did a fit  to each of the above  distributions, after rescaling  $N$ by $N/S$, where $S$ is an arbitrary scale factor ranging from from 0.001 to 2.5,  using  steps of size 0.01.

\begin{table*}
\caption{Central estimates and 1$\sigma$ bars for Messier 87 (mag) distance measurements. For the median and arithmetic mean, we have used all measurements without error bars whereas for the weighted mean we only used the ones with error bars.} 

\begin{tabular}{lccc}
\hline\hline
Central Estimate  & Median &  Arithmetic Mean & Weighted mean \\ \hline 
Tracers (with averages)  & $31.08\pm 0.09$ & $30.97\pm 0.07$ & $31.11\pm 0.008$ \\ 
Tracers (without averages) & $31.07\pm0.09$ & $30.95\pm 0.07$ & $31.10\pm 0.009$  \\  
\hline
\end{tabular}
\label{tab:1}
\end{table*}

\begin{table*}
\caption{Central estimates and 1$\sigma$ bars for Messier 87 (mag) distance measurements, using different Individual Tracers. For the median and arithmetic mean we have used all measurements without error bars whereas for the weighted mean we only used the ones with error bars.} \

\begin{tabular}{lcccc}
\hline\hline
Individual Tracers  &  Median & Arithmetic mean & Weighted mean \\ \hline 
Cepheids  & $31.02\pm 0.09$ & $31.05\pm 0.07$ & $30.96\pm 0.02$ \\ 
Planetary Nebula Luminosity Function (PNLF) & $30.86\pm 0.05$ & $30.89\pm 0.03$ & $30.91\pm 0.03$  \\ 
Surface Brightness Variation (SBF) & $31.12\pm 0.07$ & $31.10\pm 0.05$ & $31.13\pm 0.01$  \\
Tip of the Red-Gaint Branch (TRGB) & $31.05\pm 0.07$ & $31.02\pm 0.05$ & $31.03\pm 0.04$  \\
Tully-Fisher Relation (TFR) & $31.23\pm 0.08$ & $31.25\pm 0.06$ & $31.29\pm 0.02$  \\
Globular Cluster Luminosity Function (GCLF) & $31.11\pm 0.41$ & $30.74\pm 0.32$ & $31.12\pm 0.04$  \\
Supernovae (SNe) & $31.64\pm 0.11$ & $31.52\pm 0.09$ & $31.58\pm 0.06$  \\
Novae & $31.4\pm 0.08$ & $31.36\pm 0.06$ & $31.35\pm 0.14$  \\
Faber-Jackson relation & $31.14\pm 0.17$ & $31.22\pm 0.13$ & $31.02\pm 0.04$  \\
HII region & $31.2\pm 0.27$ & $31.02\pm 0.20$ & $31.27\pm 0.07$  \\
Hubble Law & $27.29\pm 0.94$ & $28.32\pm 0.70$ & $30.27\pm 0.23$  \\
Color-magnitude/luminosity & $30.84\pm 0.12$ & $30.88\pm 0.09$ & $31.02\pm 0.11$ \\
Group membership & $30.5\pm 0.10$ & $30.54\pm 0.07$ & $30.64\pm 0.12$  \\
Other methods & $30.9\pm 0.14$ & $30.92\pm 0.10$ & $31.20\pm 0.05$  \\
Averages & $31.08\pm 0.11$ & $31.10\pm 0.08$ & $31.13\pm 0.02$  \\
\hline
\end{tabular}
\label{tab:2}
\end{table*}

In order to test the efficacy of the each of the above distributions to the residuals,  we use the one-sample unbinned Kolmogorov-Smirnov (K-S) test~\cite{astroML}.  The K-S test uses  the $D$-statistic, which measures the maximum distance between two cumulative distributions. The K-S test is agnostic to the distribution against which it  is been tested, and is independent the size  of the sample. Furthermore,  one can easily obtain the $p$-value based on  the $D$-statistic~\cite{astroML}.   Therefore, the one-sample K-S test can be used to test the  goodness of fit.

The two distributions used as input to the one-sample K-S test  are the error residual histograms and the parent PDF to which it is compared.  We now present our results  for the fits to $N_{\sigma}$ for the combined  dataset as well as  separately using each of the  tracers.

\begin{itemize}

\item \textbf{All measurements}
Our results for the goodness of fits to all the four distributions  using all the tracers are summarized in Table~\ref{tab:3} . The corresponding results for   all tracers except for the ones classified as ``averages'' can be found in Table~\ref{tab:4}.
For the data with averages (cf. Table~\ref{tab:3}), we find that for all  the four estimates, the Gaussian distribution is a very poor fit with $p$-values close to or less than  $10^{-7}$. Only if the scale factor is very much different from unity (2.3), the Gaussian distribution for the median estimate is a good fit (with $p$-value of 0.6). For the scale factor of one,  only the  Cauchy distribution shows a very good fit for the median estimate.  If we exclude the measurements tagged as ``averages'', the results are comparable as can be seen from Table~\ref{tab:4}. Hence, we conclude that the distance modulus measurements show evidence for non-Gaussianity in the residuals, when we  analyze all  the measurements. Therefore, in case we need to report a central estimate, then only the median value is the most robust, since it is not affected by non-Gaussianity of the errors~\cite{Ratratheta}.

%We now  analyze the Gaussianity in each of the measurements, after splitting the dataset according to tracers. We only consider those tracers which have more than 10 measurements.

\item \textbf{Color-magnitude/Luminosity relation}
The summary statistics after considering the data obtained using color-magnitude/luminosity relation measurements  can be found in Table~\ref{tab:5}.
We find that all the four estimates show evidence for Gaussianity for the  scale factor of unity (with $p$-values greater than 0.7). This shows that there is no evidence for systematic errors using the color-magnitude/luminosity  as tracers. However, other distributions show comparable or larger $p$-values for all the central estimates.

\item \textbf{Faber-Jackson relation}
The  corresponding results when obtaining the distance modulus using  the Faber-Jackson relation can be found in Table~\ref{tab:6}.
We find that the Gaussian distribution provides a marginal fit  for scale factors of unity for all the  central estimates with $p$-values only slightly greater than 0.05. The Cauchy distribution provides the best fit with $p$-vales close to one. The Gaussian distribution is a good fit to the residuals only for scale factors between 2.5 and 2.8.

\item \textbf{Globular Cluster Luminosity Function}
The corresponding results when obtaining the distance modulus using the globular cluster luminosity function  can be found in Table~\ref{tab:7}. We find that for the median central estimate, the symmetrized $N_{\sigma}$ is consistent with the Gaussian distribution. However, for the arithmetic and weighted mean, the Gaussian distribution is not a good fit with $p$-values only slightly greater than 0.05. The estimates based on the  median and arithmetic mean have one outlier measurement, whose  distance modulus is given by  $m-M=20.9$~\cite{vdb82}. Since this measurement has no error bars provided, it  was excluded in the weighted mean-based estimate, which explains why it mainly affects the $p$-value for the arithmetic mean.

\item \textbf{Planetary Nebula Luminosity Function}
The results using the  planetary nebula luminosity function can be found in Table~\ref{tab:8}. We find that the Gaussian distribution provides a very good fit for all estimates with $p$-values $> 0.05$. However,  for all the central estimates, Laplacian distribution provides the  best fit with a $p$-value higher than the Gaussian distribution.

\item \textbf{Surface Brightness Variations}
The results using surface brightness variations  can be found in Table~\ref{tab:9}.  We find that the Gaussian distribution is a good fit to $N_{\sigma}$ for all the four central estimates with $p$-values $> 0.3$. This shows that there are no systematic errors in the distance estimates to M87 using surface brightness variations. However the Students-$t$ distribution provides a better fit than the Gaussian distribution for all the central estimates.

\item \textbf{Supernovae}
The corresponding results using supernovae as distance indicators to M87 can be found in Table~\ref{tab:10}. We find that the Gaussian distribution is very good fit to  $N_{\sigma}$ for all the central estimates. However for the median estimate and weighted mean (without correlations), the Laplacian distribution provides a better fit than the Gaussian distribution, whereas it is comparable for the weighted mean-based estimate, which accounts for correlations.

\item \textbf{Tully-Fisher relation}
The corresponding results using Tully-Fisher based distances to M87 can be found in Table~\ref{tab:11}. We find that the Gaussian distribution is not a good fit with $p$-values equal to 0.01 for the weighted mean and for the arithmetic mean. We get a good fit to the Gaussian distribution only with scale factors $> 2$  for all the central estimates. For median and weighted mean, only the Students-$t$ distribution provides
a $p$-value $> 0.05$. Therefore, the measurements based on Tully-Fisher relation contain systematic errors.

\item \textbf{Other Methods}
The results for Gaussianity tests using an assortment of other methods can be found in Table~\ref{tab:12}. Here, the median and arithmetic mean (which  do not use the error bars) provide a good fit to the Gaussian distribution. However, the weighted means do not provide a good fit to the Gaussian distribution. However, even for the  median and arithmetic means, the other three distributions such as Laplacian, Cauchy, and Students-$t$ distributions provide a better fit than the Gaussian distribution. 

 \item \textbf{Averages}
The results for Gaussianity tests for the measurements tagged as ``averages'' can be found in Table~\ref{tab:13}. We find that the residuals using all the central estimates are not consistent with Gaussian distributions (with $p$-values $<0.05$). However, this is not surprising,  since these data themselves consist of averages obtained using the different methods.
Only the Cauchy distribution provides a good fit to the underlying residuals.

\end{itemize}

\begin{table}[h]
\caption{Probabilities from K-S test for various distributions using all measurements  (including those tagged as  ``averages'')  of M87. We have used 213 data for the median and arithmetic mean, and 173 data for the weighted mean.} 

\begin{threeparttable}[t]
\begin{tabular}{cccc} 
\hline
%&\multicolumn{2}{c}{Truncated 13}& \\ \hline  

Distribution&$S$\tnote{a}&$p$\tnote{b}&$n$\tnote{c} \\ \hline
\multicolumn{4}{l}{\textbf{Median} ($(m-M)_{med}$)} \\ 
\multicolumn{1}{l}{Gaussian}&1&$1.91\times10^{-07}$& \\
&$2.38$&$0.60$& \\
\multicolumn{1}{l}{Laplacian}&1&$0.0002$&  \\
&$2.04$&$0.95$&\\
\multicolumn{1}{l}{Cauchy}&1&$0.12$& \\
&$1.12$&$0.34$& \\
\multicolumn{1}{l}{Student's $t$}&1&$0.003$&$2$ \\
&$1.65$&$0.88$&$2$ \\
\multicolumn{4}{l}{\textbf{Weighted Mean ($(m-M)_{wm+}$)}} \\ 
\multicolumn{1}{l}{Gaussian}&1&$2.22\times10^{-07}$& \\
&$2.40$&$0.66$& \\
\multicolumn{1}{l}{Laplacian}&1&$2.90\times10^{-05}$&  \\
&$1.96$&$0.58$&\\
\multicolumn{1}{l}{Cauchy}&1&$0.03$& \\
&$1.23$&$0.27$& \\
\multicolumn{1}{l}{Student's  $t$}&1&$0.0006$&$2$ \\
&$1.66$&$0.62$&$2$ \\
\multicolumn{4}{l}{\textbf{Weighted Mean ($(m-M)_{wm-}$) }} \\ 
\multicolumn{1}{l}{Gaussian}&1&$1.38\times10^{-07}$& \\
&$2.41$&$0.67$& \\
\multicolumn{1}{l}{Laplacian}&1&$2.14\times10^{-05}$&  \\
&$2.01$&$0.63$&\\
\multicolumn{1}{l}{Cauchy}&1&$0.02$& \\
&$1.26$&$0.30$& \\
\multicolumn{1}{l}{Student's $t$}&1&$0.0004$&$2$ \\
&$1.70$&$0.67$&$2$ \\
\multicolumn{4}{l}{\textbf{Arithmetic Mean ($(m-M)_{m}$)}} \\ 
\multicolumn{1}{l}{Gaussian}&1&$3.01\times10^{-15}$& \\
&$2.17$&$0.003$& \\
\multicolumn{1}{l}{Laplacian}&1&$1.13\times10^{-11}$&  \\
&$2.33$&$0.002$&\\
\multicolumn{1}{l}{Cauchy}&1&$2.43\times10^{-06}$& \\
&$1.62$&$0.002$& \\
\multicolumn{1}{l}{Student's  $t$}&1&$2.14\times10^{-09}$&$2$ \\
&$1.87$&$0.003$&$2$ \\
\hline
\end{tabular}

\begin{tablenotes}
\item[a] \footnotesize{The scale factor (other than 1)  which maximizes $p$}
\item[b] \footnotesize{$p$-value that the data is derived from the PDF}
\item[c] \footnotesize{The value $n$ in the students $t$-distribution}
\end{tablenotes}

\end{threeparttable}
\label{tab:3}
\end{table}

\begin{table}[h]
\caption{Probabilities from K-S test for various distributions using all measurements (except those tagged as  averages) measurement data of M87. We have used 190 data points for the median and arithmetic mean, and 153 data for the weighted mean. All variables have the same meaning  as  in Table~\ref{tab:3}.}\

\begin{threeparttable}[t]
\begin{tabular}{cccc} 
\hline
%&\multicolumn{2}{c}{Truncated 13}& \\ \hline  

Distribution&$S$\tnote{a}&$p$\tnote{b}&$n$\tnote{c} \\ 
\hline
\multicolumn{4}{l}{\textbf{Median} ($(m-M)_{med}$)} \\ 
\multicolumn{1}{l}{Gaussian}&1&$1.40\times10^{-05}$& \\
&$2.19$&$0.65$& \\
\multicolumn{1}{l}{Laplacian}&1&$0.001$&  \\
&$1.87$&$0.93$&\\
\multicolumn{1}{l}{Cauchy}&1&$0.22$& \\
&$1.09$&$0.40$& \\
\multicolumn{1}{l}{Student's $t$}&1&$0.02$&$2$ \\
&$1.54$&$0.95$&$2$ \\
\multicolumn{4}{l}{\textbf{Weighted Mean ($(m-M)_{wm+}$)}} \\ 
\multicolumn{1}{l}{Gaussian}&1&$4.90\times10^{-06}$& \\
&$2.21$&$0.58$& \\
\multicolumn{1}{l}{Laplacian}&1&$0.0004$&  \\
&$1.83$&$0.63$&\\
\multicolumn{1}{l}{Cauchy}&1&$0.09$& \\
&$1.14$&$0.30$& \\
\multicolumn{1}{l}{Student's  $t$}&1&$0.005$&$2$ \\
&$1.54$&$0.67$&$2$ \\
\multicolumn{4}{l}{\textbf{Weighted Mean ($(m-M)_{wm-}$) }} \\ 
\multicolumn{1}{l}{Gaussian}&1&$1.70\times10^{-06}$& \\
&$2.19$&$0.53$& \\
\multicolumn{1}{l}{Laplacian}&1&$0.0003$&  \\
&$1.82$&$0.64$&\\
\multicolumn{1}{l}{Cauchy}&1&$0.13$& \\
&$1.12$&$0.33$& \\
\multicolumn{1}{l}{Student's $t$}&1&$0.003$&$2$ \\
&$1.56$&$0.63$&$2$ \\
\multicolumn{4}{l}{\textbf{Arithmetic Mean ($(m-M)_{m}$)}} \\ 
\multicolumn{1}{l}{Gaussian}&1&$4.85\times10^{-13}$& \\
&$1.98$&$0.001$& \\
\multicolumn{1}{l}{Laplacian}&1&$1.57\times10^{-10}$&  \\
&$2.06$&$0.001$&\\
\multicolumn{1}{l}{Cauchy}&1&$4.46\times10^{-06}$& \\
&$1.43$&$0.001$& \\
\multicolumn{1}{l}{Student's  $t$}&1&$2.85\times10^{-08}$&$2$ \\
&$1.66$&$0.002$&$2$ \\
\hline
\end{tabular}

\end{threeparttable}
\label{tab:4}
\end{table}

\begin{table}[h]
\caption{Probabilities from K-S test for various distributions using the Color-magnitude/luminosity relation measurement data of M87. We have used 11 data for the median and arithmetic mean, and seven measurements for Weighted mean. All variables have the same meaning  as  in Table~\ref{tab:3}.}\

\begin{threeparttable}[t]
\begin{tabular}{cccc} 
\hline
%&\multicolumn{2}{c}{Truncated 13}& \\ \hline  

Distribution&$S$\tnote{a}&$p$\tnote{b}&$n$\tnote{c} \\ 
\hline
\multicolumn{4}{l}{\textbf{Median} ($(m-M)_{med}$)} \\ 
\multicolumn{1}{l}{Gaussian}&1&$0.93$& \\
&$0.98$&$0.94$& \\
\multicolumn{1}{l}{Laplacian}&1&$0.95$&  \\
&$0.86$&$0.99$&\\
\multicolumn{1}{l}{Cauchy}&1&$0.71$& \\
&$0.56$&$0.99$& \\
\multicolumn{1}{l}{Student's $t$}&1&$0.84$&$2$ \\
&$0.70$&$0.99$&$2$ \\
\multicolumn{4}{l}{\textbf{Weighted Mean ($(m-M)_{wm+}$)}} \\ 
\multicolumn{1}{l}{Gaussian}&1&$0.59$& \\
&$1.14$&$0.70$& \\
\multicolumn{1}{l}{Laplacian}&1&$0.58$&  \\
&$1.23$&$0.72$&\\
\multicolumn{1}{l}{Cauchy}&1&$0.64$& \\
&$0.86$&$0.71$& \\
\multicolumn{1}{l}{Student's  $t$}&1&$0.69$&$2$ \\
&$0.98$&$0.71$&$2$ \\
\multicolumn{4}{l}{\textbf{Weighted Mean ($(m-M)_{wm-}$) }} \\ 
\multicolumn{1}{l}{Gaussian}&1&$0.48$& \\
&$1.24$&$0.66$& \\
\multicolumn{1}{l}{Laplacian}&1&$0.48$&  \\
&$1.34$&$0.68$&\\
\multicolumn{1}{l}{Cauchy}&1&$0.64$& \\
&$0.93$&$0.67$& \\
\multicolumn{1}{l}{Student's $t$}&1&$0.61$&$2$ \\
&$1.07$&$0.67$&$2$ \\
\multicolumn{4}{l}{\textbf{Arithmetic Mean ($(m-M)_{m}$)}} \\ 
\multicolumn{1}{l}{Gaussian}&1&$0.79$& \\
&$1.09$&$0.84$& \\
\multicolumn{1}{l}{Laplacian}&1&$0.73$&  \\
&$1.19$&$0.84$&\\
\multicolumn{1}{l}{Cauchy}&1&$0.73$& \\
&$0.83$&$0.84$& \\
\multicolumn{1}{l}{Student's  $t$}&1&$0.81$&$2$ \\
&$0.95$&$0.84$&$2$ \\
\hline
\end{tabular}

\end{threeparttable}
\label{tab:5}
\end{table}

.

\begin{table}[h]
\caption{Probabilities from K-S test for various distributions using the Faber-Jackson relation measurement data of M87. We have used 11 data for median and arithmetic mean, and 10 for the weighted mean. All variables have the same meaning  as  in Table~\ref{tab:3}.}\

\begin{threeparttable}[t]
\begin{tabular}{cccc} 
\hline
%&\multicolumn{2}{c}{Truncated 13}& \\ \hline  

Distribution&$S$ &$p$ &$n$ \\ 
\hline
\multicolumn{4}{l}{\textbf{Median} ($(m-M)_{med}$)} \\ 
\multicolumn{1}{l}{Gaussian}&1&$0.06$& \\
&$2.50$&$0.84$& \\
\multicolumn{1}{l}{Laplacian}&1&$0.13$&  \\
&$2.35$&$0.79$&\\
\multicolumn{1}{l}{Cauchy}&1&$0.37$& \\
&$1.59$&$0.76$& \\
\multicolumn{1}{l}{Student's $t$}&1&$0.19$&$2$ \\
&$2.00$&$0.80$&$2$ \\
\multicolumn{4}{l}{\textbf{Weighted Mean ($(m-M)_{wm+}$)}} \\ 
\multicolumn{1}{l}{Gaussian}&1&$0.08$& \\
&$2.68$&$0.84$& \\
\multicolumn{1}{l}{Laplacian}&1&$0.17$&  \\
&$2.46$&$0.82$&\\
\multicolumn{1}{l}{Cauchy}&1&$0.46$& \\
&$1.63$&$0.81$& \\
\multicolumn{1}{l}{Student's  $t$}&1&$0.23$&$2$ \\
&$2.12$&$0.82$&$2$ \\
\multicolumn{4}{l}{\textbf{Weighted Mean ($(m-M)_{wm-}$) }} \\ 
\multicolumn{1}{l}{Gaussian}&1&$0.08$& \\
&$2.84$&$0.90$& \\
\multicolumn{1}{l}{Laplacian}&1&$0.16$&  \\
&$2.87$&$0.93$&\\
\multicolumn{1}{l}{Cauchy}&1&$0.44$& \\
&$1.99$&$0.93$& \\
\multicolumn{1}{l}{Student's $t$}&1&$0.22$&$2$ \\
&$2.38$&$0.91$&$2$ \\
\multicolumn{4}{l}{\textbf{Arithmetic Mean ($(m-M)_{m}$)}} \\ 
\multicolumn{1}{l}{Gaussian}&1&$0.10$& \\
&$2.52$&$0.95$& \\
\multicolumn{1}{l}{Laplacian}&1&$0.19$&  \\
&$2.36$&$0.90$&\\
\multicolumn{1}{l}{Cauchy}&1&$0.48$& \\
&$1.57$&$0.86$& \\
\multicolumn{1}{l}{Student's  $t$}&1&$0.27$&$2$ \\
&$2.00$&$0.91$&$2$ \\
\hline
\end{tabular}

\end{threeparttable}
\label{tab:6}
\end{table}

\begin{table}[h]
\caption{Probabilities from K-S test for various distributions using the Globular Cluster Luminosity Function (GCLF) measurement data of M87. We have used 32 data for the median and arithmetic mean, and 23 for the weighted mean. All variables have the same meaning  as  in Table~\ref{tab:3}.}\

\begin{threeparttable}[t]
\begin{tabular}{cccc} 
\hline
%&\multicolumn{2}{c}{Truncated 13}& \\ \hline  

Distribution&$S$\tnote{a}&$p$\tnote{b}&$n$\tnote{c} \\ 
\hline
\multicolumn{4}{l}{\textbf{Median} ($(m-M)_{med}$)} \\ 
\multicolumn{1}{l}{Gaussian}&1&$0.97$& \\
&$0.99$&$0.98$& \\
\multicolumn{1}{l}{Laplacian}&1&$0.88$&  \\
&$0.94$&$0.94$&\\
\multicolumn{1}{l}{Cauchy}&1&$0.25$& \\
&$0.58$&$0.81$& \\
\multicolumn{1}{l}{Student's $t$}&1&$0.63$&$2$ \\
&$0.80$&$0.94$&$2$ \\
\multicolumn{4}{l}{\textbf{Weighted Mean ($(m-M)_{wm+}$)}} \\ 
\multicolumn{1}{l}{Gaussian}&1&$0.06$& \\
&$2.17$&$0.89$& \\
\multicolumn{1}{l}{Laplacian}&1&$0.11$&  \\
&$2.37$&$0.89$&\\
\multicolumn{1}{l}{Cauchy}&1&$0.37$& \\
&$1.56$&$0.84$& \\
\multicolumn{1}{l}{Student's  $t$}&1&$0.21$&$2$ \\
&$1.89$&$0.89$&$2$ \\
\multicolumn{4}{l}{\textbf{Weighted Mean ($(m-M)_{wm-}$) }} \\ 
\multicolumn{1}{l}{Gaussian}&1&$0.05$& \\
&$2.23$&$0.89$& \\
\multicolumn{1}{l}{Laplacian}&1&$0.10$&  \\
&$2.42$&$0.89$&\\
\multicolumn{1}{l}{Cauchy}&1&$0.35$& \\
&$1.59$&$0.84$& \\
\multicolumn{1}{l}{Student's $t$}&1&$0.19$&$2$ \\
&$1.94$&$0.89$&$2$ \\
\multicolumn{4}{l}{\textbf{Arithmetic Mean ($(m-M)_{m}$)}} \\ 
\multicolumn{1}{l}{Gaussian}&1&$0.001$& \\
&$1.30$&$0.01$& \\
\multicolumn{1}{l}{Laplacian}&1&$0.002$&  \\
&$1.35$&$0.01$&\\
\multicolumn{1}{l}{Cauchy}&1&$0.01$& \\
&$0.92$&$0.02$& \\
\multicolumn{1}{l}{Student's  $t$}&1&$0.006$&$2$ \\
&$1.09$&$0.01$&$2$ \\
\hline
\end{tabular}

\end{threeparttable}
\label{tab:7}
\end{table}

\begin{table}[h]
\caption{Probabilities from K-S test for various distributions using the Planetary Nebula Luminosity Function (PNLF) measurement data of M87. We have used 12 data for the median and arithmetic mean, and 10 for the  weighted mean. All variables have the same meaning  as  in Table~\ref{tab:3}.}\

\begin{threeparttable}[t]
\begin{tabular}{cccc} 
\hline
%&\multicolumn{2}{c}{Truncated 13}& \\ \hline  

Distribution&$S$\tnote{a}&$p$\tnote{b}&$n$\tnote{c} \\ 
\hline
\multicolumn{4}{l}{\textbf{Median} ($(m-M)_{med}$)} \\ 
\multicolumn{1}{l}{Gaussian}&1&$0.75$& \\
&$0.51$&$0.98$& \\
\multicolumn{1}{l}{Laplacian}&1&$0.81$&  \\
&$0.50$&$0.99$&\\
\multicolumn{1}{l}{Cauchy}&1&$0.53$& \\
&$0.35$&$0.99$& \\
\multicolumn{1}{l}{Student's $t$}&1&$0.69$&$2$ \\
&$0.42$&$0.99$&$2$ \\
\multicolumn{4}{l}{\textbf{Weighted Mean ($(m-M)_{wm+}$)}} \\ 
\multicolumn{1}{l}{Gaussian}&1&$0.10$& \\
&$0.67$&$0.13$& \\
\multicolumn{1}{l}{Laplacian}&1&$0.11$&  \\
&$0.80$&$0.13$&\\
\multicolumn{1}{l}{Cauchy}&1&$0.09$& \\
&$0.53$&$0.13$& \\
\multicolumn{1}{l}{Student's  $t$}&1&$0.09$&$2$ \\
&$0.59$&$0.13$&$2$ \\
\multicolumn{4}{l}{\textbf{Weighted Mean ($(m-M)_{wm-}$) }} \\ 
\multicolumn{1}{l}{Gaussian}&1&$0.10$& \\
&$0.69$&$0.13$& \\
\multicolumn{1}{l}{Laplacian}&1&$0.11$&  \\
&$0.83$&$0.13$&\\
\multicolumn{1}{l}{Cauchy}&1&$0.09$& \\
&$0.55$&$0.13$& \\
\multicolumn{1}{l}{Student's $t$}&1&$0.09$&$2$ \\
&$0.61$&$0.13$&$2$ \\
\multicolumn{4}{l}{\textbf{Arithmetic Mean ($(m-M)_{m}$)}} \\ 
\multicolumn{1}{l}{Gaussian}&1&$0.15$& \\
&$0.53$&$0.50$& \\
\multicolumn{1}{l}{Laplacian}&1&$0.18$&  \\
&$0.52$&$0.53$&\\
\multicolumn{1}{l}{Cauchy}&1&$0.10$& \\
&$0.37$&$0.53$& \\
\multicolumn{1}{l}{Student's  $t$}&1&$0.12$&$2$ \\
&$0.44$&$0.52$&$2$ \\
\hline
\end{tabular}

\end{threeparttable}
\label{tab:8}
\end{table}

\begin{table}[h]
\caption{Probabilities from K-S test for various distributions using the Surface Brightness Variations (SBF) measurement data of M87. We have used 18 data for the median and arithmetic mean, and 17 for the weighted mean. All variables have the same meaning  as  in Table~\ref{tab:3}.}

\begin{threeparttable}[t]
\begin{tabular}{cccc} 
\hline
%&\multicolumn{2}{c}{Truncated 13}& \\ \hline  

Distribution&$S$\tnote{a}&$p$\tnote{b}&$n$\tnote{c} \\ 
\hline
\multicolumn{4}{l}{\textbf{Median} ($(m-M)_{med}$)} \\ 
\multicolumn{1}{l}{Gaussian}&1&$0.53$& \\
&$1.23$&$0.80$& \\
\multicolumn{1}{l}{Laplacian}&1&$0.80$&  \\
&$1.14$&$0.90$&\\
\multicolumn{1}{l}{Cauchy}&1&$0.65$& \\
&$0.78$&$0.88$& \\
\multicolumn{1}{l}{Student's $t$}&1&$0.85$&$2$ \\
&$0.94$&$0.90$&$2$ \\
\multicolumn{4}{l}{\textbf{Weighted Mean ($(m-M)_{wm+}$)}} \\ 
\multicolumn{1}{l}{Gaussian}&1&$0.32$& \\
&$1.41$&$0.68$& \\
\multicolumn{1}{l}{Laplacian}&1&$0.35$&  \\
&$1.44$&$0.73$&\\
\multicolumn{1}{l}{Cauchy}&1&$0.71$& \\
&$1.01$&$0.73$& \\
\multicolumn{1}{l}{Student's  $t$}&1&$0.53$&$2$ \\
&$1.17$&$0.72$&$2$ \\
\multicolumn{4}{l}{\textbf{Weighted Mean ($(m-M)_{wm-}$) }} \\ 
\multicolumn{1}{l}{Gaussian}&1&$0.31$& \\
&$1.51$&$0.66$& \\
\multicolumn{1}{l}{Laplacian}&1&$0.34$&  \\
&$1.47$&$0.75$&\\
\multicolumn{1}{l}{Cauchy}&1&$0.71$& \\
&$1.04$&$0.74$& \\
\multicolumn{1}{l}{Student's $t$}&1&$0.52$&$2$ \\
&$1.20$&$0.74$&$2$ \\
\multicolumn{4}{l}{\textbf{Arithmetic Mean ($(m-M)_{m}$)}} \\ 
\multicolumn{1}{l}{Gaussian}&1&$0.43$& \\
&$1.53$&$0.96$& \\
\multicolumn{1}{l}{Laplacian}&1&$0.75$&  \\
&$1.37$&$0.97$&\\
\multicolumn{1}{l}{Cauchy}&1&$0.86$& \\
&$0.85$&$0.94$& \\
\multicolumn{1}{l}{Student's  $t$}&1&$0.89$&$2$ \\
&$1.17$&$0.98$&$2$ \\
\hline
\end{tabular}

\end{threeparttable}
\label{tab:9}
\end{table}

\begin{table}[h]
\caption{Probabilities from the K-S test for various distributions using the Supernovae (SNe) I, Ia, II  measurement data of M87. We have used 18 data for the median and arithmetic mean, and 16 for the weighted mean. All variables have the same meaning  as  in Table~\ref{tab:3}.}\ 

\begin{threeparttable}[t]
\begin{tabular}{cccc} 
\hline
%&\multicolumn{2}{c}{Truncated 13}& \\ \hline  

Distribution&$S$\tnote{a}&$p$\tnote{b}&$n$\tnote{c} \\ 
\hline
\multicolumn{4}{l}{\textbf{Median} ($(m-M)_{med}$)} \\ 
\multicolumn{1}{l}{Gaussian}&1&$0.41$& \\
&$1.06$&$0.46$& \\
\multicolumn{1}{l}{Laplacian}&1&$0.42$&  \\
&$0.87$&$0.60$&\\
\multicolumn{1}{l}{Cauchy}&1&$0.14$& \\
&$0.47$&$0.76$& \\
\multicolumn{1}{l}{Student's $t$}&1&$0.27$&$2$ \\
&$0.73$&$0.64$&$2$ \\
\multicolumn{4}{l}{\textbf{Weighted Mean ($(m-M)_{wm+}$)}} \\ 
\multicolumn{1}{l}{Gaussian}&1&$0.30$& \\
&$0.85$&$0.42$& \\
\multicolumn{1}{l}{Laplacian}&1&$0.35$&  \\
&$0.92$&$0.41$&\\
\multicolumn{1}{l}{Cauchy}&1&$0.18$& \\
&$0.64$&$0.42$& \\
\multicolumn{1}{l}{Student's  $t$}&1&$0.23$&$2$ \\
&$0.73$&$0.42$&$2$ \\
\multicolumn{4}{l}{\textbf{Weighted Mean ($(m-M)_{wm-}$) }} \\ 
\multicolumn{1}{l}{Gaussian}&1&$0.41$& \\
&$0.99$&$0.42$& \\
\multicolumn{1}{l}{Laplacian}&1&$0.40$&  \\
&$1.07$&$0.41$&\\
\multicolumn{1}{l}{Cauchy}&1&$0.24$& \\
&$0.75$&$0.42$& \\
\multicolumn{1}{l}{Student's $t$}&1&$0.31$&$2$ \\
&$0.86$&$0.42$&$2$ \\
\multicolumn{4}{l}{\textbf{Arithmetic Mean ($(m-M)_{m}$)}} \\ 
\multicolumn{1}{l}{Gaussian}&1&$0.29$& \\
&$1.14$&$0.35$& \\
\multicolumn{1}{l}{Laplacian}&1&$0.23$&  \\
&$1.30$&$0.36$&\\
\multicolumn{1}{l}{Cauchy}&1&$0.31$& \\
&$0.89$&$0.35$& \\
\multicolumn{1}{l}{Student's  $t$}&1&$0.35$&$2$ \\
&$0.99$&$0.36$&$2$ \\
\hline
\end{tabular}

\end{threeparttable}
\label{tab:10}
\end{table}

\begin{table}[h]
\caption{Probabilities from K-S test for various distributions using the Tully-Fisher Relations (TFR) measurement data of M87. We have used 36 data for the median and arithmetic mean, and 32 for the weighted mean. All variables have the same meaning  as  in Table~\ref{tab:3}.}

\begin{threeparttable}[t]
\begin{tabular}{cccc} 
\hline
%&\multicolumn{2}{c}{Truncated 13}& \\ \hline  

Distribution&$S$\tnote{a}&$p$\tnote{b}&$n$\tnote{c} \\ 
\hline
\multicolumn{4}{l}{\textbf{Median} ($(m-M)_{med}$)} \\ 
\multicolumn{1}{l}{Gaussian}&1&$0.01$& \\
&$2.35$&$0.88$& \\
\multicolumn{1}{l}{Laplacian}&1&$0.05$&  \\
&$2.21$&$0.80$&\\
\multicolumn{1}{l}{Cauchy}&1&$0.28$& \\
&$1.49$&$0.76$& \\
\multicolumn{1}{l}{Student's $t$}&1&$0.11$&$2$ \\
&$1.82$&$0.86$&$2$ \\
\multicolumn{4}{l}{\textbf{Weighted Mean ($(m-M)_{wm+}$)}} \\ 
\multicolumn{1}{l}{Gaussian}&1&$0.01$& \\
&$2.90$&$0.99$& \\
\multicolumn{1}{l}{Laplacian}&1&$0.04$&  \\
&$2.57$&$0.97$&\\
\multicolumn{1}{l}{Cauchy}&1&$0.27$& \\
&$1.60$&$0.89$& \\
\multicolumn{1}{l}{Student's  $t$}&1&$0.09$&$2$ \\
&$2.14$&$0.98$&$2$ \\
\multicolumn{4}{l}{\textbf{Weighted Mean ($(m-M)_{wm-}$) }} \\ 
\multicolumn{1}{l}{Gaussian}&1&$0.01$& \\
&$2.99$&$0.99$& \\
\multicolumn{1}{l}{Laplacian}&1&$0.04$&  \\
&$2.70$&$0.98$&\\
\multicolumn{1}{l}{Cauchy}&1&$0.26$& \\
&$1.65$&$0.91$& \\
\multicolumn{1}{l}{Student's $t$}&1&$0.09$&$2$ \\
&$2.24$&$0.99$&$2$ \\
\multicolumn{4}{l}{\textbf{Arithmetic Mean ($(m-M)_{m}$)}} \\ 
\multicolumn{1}{l}{Gaussian}&1&$0.01$& \\
&$2.58$&$0.99$& \\
\multicolumn{1}{l}{Laplacian}&1&$0.03$&  \\
&$2.43$&$0.93$&\\
\multicolumn{1}{l}{Cauchy}&1&$0.24$& \\
&$1.54$&$0.83$& \\
\multicolumn{1}{l}{Student's  $t$}&1&$0.08$&$2$ \\
&$2.05$&$0.96$&$2$ \\
\hline
\end{tabular}

\end{threeparttable}
\label{tab:11}
\end{table}

\begin{table}[h]
\caption{Probabilities from K-S test for various distributions using the measurements classified as ``Other methods''. We have used 15 data for the median and arithmetic mean, and 13 for weighted mean. All variables have the same meaning  as  in Table~\ref{tab:3}.}
\begin{threeparttable}[t]
\begin{tabular}{cccc} 
\hline
%&\multicolumn{2}{c}{Truncated 13}& \\ \hline  

Distribution&$S$\tnote{a}&$p$\tnote{b}&$n$\tnote{c} \\ 
\hline
\multicolumn{4}{l}{\textbf{Median} ($(m-M)_{med}$)} \\ 
\multicolumn{1}{l}{Gaussian}&1&$0.54$& \\
&$1.66$&$0.85$& \\
\multicolumn{1}{l}{Laplacian}&1&$0.74$&  \\
&$1.46$&$0.96$&\\
\multicolumn{1}{l}{Cauchy}&1&$0.97$& \\
&$0.97$&$0.98$& \\
\multicolumn{1}{l}{Student's $t$}&1&$0.88$&$2$ \\
&$1.21$&$0.96$&$2$ \\
\multicolumn{4}{l}{\textbf{Weighted Mean ($(m-M)_{wm+}$)}} \\ 
\multicolumn{1}{l}{Gaussian}&1&$0.03$& \\
&$6.80$&$0.06$& \\
\multicolumn{1}{l}{Laplacian}&1&$0.02$&  \\
&$7.42$&$0.05$&\\
\multicolumn{1}{l}{Cauchy}&1&$0.03$& \\
&$5.17$&$0.06$& \\
\multicolumn{1}{l}{Student's  $t$}&1&$0.03$&$2$ \\
&$5.90$&$0.06$&$2$ \\
\multicolumn{4}{l}{\textbf{Weighted Mean ($(m-M)_{wm-}$) }} \\ 
\multicolumn{1}{l}{Gaussian}&1&$0.03$& \\
&$7.90$&$0.06$& \\
\multicolumn{1}{l}{Laplacian}&1&$0.02$&  \\
&$8.61$&$0.06$&\\
\multicolumn{1}{l}{Cauchy}&1&$0.03$& \\
&$5.99$&$0.60$& \\
\multicolumn{1}{l}{Student's $t$}&1&$0.03$&$2$ \\
&$6.86$&$0.06$&$2$ \\
\multicolumn{4}{l}{\textbf{Arithmetic Mean ($(m-M)_{m}$)}} \\ 
\multicolumn{1}{l}{Gaussian}&1&$0.52$& \\
&$2.01$&$0.84$& \\
\multicolumn{1}{l}{Laplacian}&1&$0.64$&  \\
&$1.68$&$0.93$&\\
\multicolumn{1}{l}{Cauchy}&1&$0.97$& \\
&$0.97$&$0.98$& \\
\multicolumn{1}{l}{Student's  $t$}&1&$0.78$&$2$ \\
&$1.43$&$0.94$&$2$ \\
\hline
\end{tabular}

\end{threeparttable}
\label{tab:12}
\end{table}

\begin{table}[h]
\caption{Probabilities from K-S test for various distributions using the M87 measurements tagged as ``averages''. We have used 21 measurements for the median and arithmetic mean, and 18 for the weighted mean. All variables have the same meaning  as  in Table~\ref{tab:3}.}

\begin{threeparttable}[t]
\begin{tabular}{cccc} 
\hline
%&\multicolumn{2}{c}{Truncated 13}& \\ \hline  

Distribution&$S$\tnote{a}&$p$\tnote{b}&$n$\tnote{c} \\ 
\hline
\multicolumn{4}{l}{\textbf{Median} ($(m-M)_{med}$)} \\ 
\multicolumn{1}{l}{Gaussian}&1&$0.01$& \\
&$2.46$&$0.84$& \\
\multicolumn{1}{l}{Laplacian}&1&$0.05$&  \\
&$2.50$&$0.88$&\\
\multicolumn{1}{l}{Cauchy}&1&$0.29$& \\
&$1.72$&$0.87$& \\
\multicolumn{1}{l}{Student's $t$}&1&$0.10$&$2$ \\
&$2.06$&$0.86$&$2$ \\
\multicolumn{4}{l}{\textbf{Weighted Mean ($(m-M)_{wm+}$)}} \\ 
\multicolumn{1}{l}{Gaussian}&1&$0.02$& \\
&$3.30$&$0.99$& \\
\multicolumn{1}{l}{Laplacian}&1&$0.07$&  \\
&$3.23$&$0.99$&\\
\multicolumn{1}{l}{Cauchy}&1&$0.32$& \\
&$2.12$&$0.98$& \\
\multicolumn{1}{l}{Student's  $t$}&1&$0.12$&$2$ \\
&$2.70$&$0.99$&$2$ \\
\multicolumn{4}{l}{\textbf{Weighted Mean ($(m-M)_{wm-}$) }} \\ 
\multicolumn{1}{l}{Gaussian}&1&$0.01$& \\
&$3.77$&$0.99$& \\
\multicolumn{1}{l}{Laplacian}&1&$0.04$&  \\
&$3.64$&$0.99$&\\
\multicolumn{1}{l}{Cauchy}&1&$0.22$& \\
&$2.37$&$0.97$& \\
\multicolumn{1}{l}{Student's $t$}&1&$0.07$&$2$ \\
&$3.08$&$0.99$&$2$ \\
\multicolumn{4}{l}{\textbf{Arithmetic Mean ($(m-M)_{m}$)}} \\ 
\multicolumn{1}{l}{Gaussian}&1&$0.02$& \\
&$2.90$&$0.99$& \\
\multicolumn{1}{l}{Laplacian}&1&$0.06$&  \\
&$2.91$&$0.99$&\\
\multicolumn{1}{l}{Cauchy}&1&$0.32$& \\
&$1.80$&$0.93$& \\
\multicolumn{1}{l}{Student's  $t$}&1&$0.11$&$2$ \\
&$2.44$&$0.99$&$2$ \\
\hline
\end{tabular}

\end{threeparttable}
\label{tab:13}
\end{table}

\iffalse

\fi

\section{Conclusions}
\label{sec:conclusions}

Recently, D20 did an extensive data mining of literature  to compile all the distance measurements  to M87 using the Galactic center, LMC and M31 as distance anchors. They also classified all measurements into 15 distinct tracers, of which eight tracers contained more than  10 measurements.  We carried out an extensive meta-analysis for all these measurements along the same lines as our previous works~\cite{Rajan,Rajan2,Bethapudi}, which follow in spirit similar work done by Ratra et al~\cite{RatraD2,RatraGC,Ratratheta} (and references therein). The main goal was to characterize the Gaussianity  in the error residuals of these measurements, when using the full dataset as well as after classifying them according to the type of tracers used. Any evidence for non-Gaussianity in the residuals would point to   systematic errors  in these measurements~\cite{Gott}. Therefore, our work complements the extensive analysis carried out in D20.

 For this purpose, we calculated the central estimate   using both the weighted mean (with and without correlations), arithmetic mean  as well as the median value. The median estimate does not incorporate any errors. This was done for the full dataset and also after classifying the measurements according to the type of tracers used as long as each tracer contained more than 10 measurements. These results can be found in Table~\ref{tab:1} and Table~\ref{tab:2} respectively.
  We then fit these residuals to four distributions, viz. Gaussian, Laplace, Cauchy, and Student's $t$ distribution using the one-sample K-S test. These results can be found in Tables~\ref{tab:3}, ~\ref{tab:4}, ~\ref{tab:5}, ~\ref{tab:6}, ~\ref{tab:7}, ~\ref{tab:8}, ~\ref{tab:9},  ~\ref{tab:10}, ~\ref{tab:11} ~\ref{tab:12}, and ~\ref{tab:13}.

Our conclusions are as follows:
\begin{itemize}
\item The central estimates which we obtained using all the three central estimates agree with the estimates in D20 to within $1\sigma$.
\item If we look at the measurements after classifying them according to tracers, except for Hubble law, all the measurements are consistent with each other. The measurements based on Hubble's law are inconsistent to within $3-4\sigma$. 
\item When we consider the full dataset, the residuals using the weighted mean are a poor fit to  the residuals. Therefore the median estimate which  we obtain ($31.08 \pm 0.09$) should be used as the central estimate.
\item We find that after splitting the data according to the tracers,   the measurements based on the Tully-Fisher relation and those tagged as ``Averages'' show a poor fit to the Gaussian distribution for all the central estimates. A good fit to Gaussian distribution is only obtained for scale factors between 2.5 and 3.8. This indicates that these measurements contain unaccounted for systematic errors.
\item The residuals using the measurements based on the Faber-Jackson relation are only marginally consistent with the Gaussian distribution (for all estimates) with $p$-values between 0.05-0.1.
\item For globular cluster luminosity function based measurements as well as those classified as ``Other'', only the residuals using median estimate show a good fit to  Gaussian distribution. All other estimates have a poor fit to the Gaussian distribution.
\item For all other measurements classified according to tracers, the residuals  are consistent with a Gaussian distribution.  However, other distributions such as Laplace or Cauchy also provide an equally good or better fit to the residuals.

\end{itemize}

\textbf{Note added:} \rthis{After this work was submitted, we were informed that another work on similar lines was under preparation, and has been submitted for publication at the time of writing~\cite{Ratra23}. This work  focuses on using the median estimates to estimate the systematic errors in the distance measurements, whereas  the emphasis in our work was on testing the Gaussianity of the error residuals.}

\begin{acknowledgements}
G. Ramakrishnan was supported by a SURE internship at IIT Hyderabad. \rthis{We are grateful to one of the referees Bharat Ratra for useful feedback on our manuscript and sharing the results of ~\cite{Ratra23}.}
\end{acknowledgements}
\bibliography{main}
\end{document}